\documentclass[preprint,12pt]{elsarticle}

\usepackage{amssymb}
\usepackage{amsmath}
\usepackage{float}
\usepackage{subcaption}
\usepackage{graphicx} 
\usepackage{tikz}
\usetikzlibrary{arrows.meta, positioning, shapes.geometric, calc}

\journal{Artificial Intelligence in Medicine}

\begin{document}

\begin{frontmatter}



\title{Physics Informed Neural Network model for the dynamical study of Abdominal Aortic Aneurysm}


\author{Adrián Robles Arques, Martín Ruiz Fernandez, Javier Sanchis, Miguel A. Teruel, Juan Trujillo} 


\affiliation{organization={Lucentia Research; Instituto Universitario de Investigación en Informática, Universidad de Alicante},
            addressline={Carr. San Vicent del Raspeig}, 
            city={Alicante},
            postcode={03690}, 
            state={Comunidad Valenciana},
            country={España}}

\begin{abstract}
We present the development and application of a three-dimensional Physics-Informed Neural Network (PINN) framework for the investigation of haemodynamic behaviour in the human aorta. The model incorporates a time-resolved simulation of pulsatile blood flow over a two-minute interval, enabling the extraction of pressure and velocity fields with high temporal fidelity. The mechanical stress exerted on the aortic wall was quantified through Laplace’s law, with temporal averaging applied to derive representative stress distributions. This approach circumvents the computational overhead associated with conventional computational fluid dynamics (CFD) methods by eliminating mesh generation and exploiting the automatic differentiation capabilities inherent to neural networks. The proposed methodology demonstrates that PINNs can serve as an efficient and accurate alternative for modelling complex vascular flow phenomena, offering significant advantages in scalability and computational cost reduction while maintaining physical consistency.
\end{abstract}











\end{frontmatter}



\section{Introduction} \label{sec:intro}
\label{sec1}

The simulation of fluid dynamics constitutes a central pillar in both scientific research and engineering practice. Accurate modelling of blood flow, aerodynamics, and industrial transport processes has enabled significant advances in medicine, aerospace, and energy systems. At the heart of these simulations lies the solution of partial differential equations (PDEs), which govern the conservation of mass, momentum, and energy in fluid dynamics. Classical numerical approaches, such as finite element and finite volume methods, have long provided robust frameworks for solving these equations. Their success is evident in the widespread adoption of computational fluid dynamics (CFD) across multiple disciplines, including haemodynamic simulation. \cite{BlodFlowCFDAAA, HeartBloodFlowCFD} 

Despite their utility, conventional CFD methods encounter notable limitations when applied to complex flow regimes. High‑fidelity simulations often demand extremely fine mesh resolution, leading to high computational costs \cite{CFDBloodFlowLim}. Furthermore, intricate boundary conditions, non-linearities, and stiff problems challenge the stability and convergence of traditional solvers. \cite{SanchisCFD_1, SanchisCFD_2} These constraints are particularly pronounced in biomedical applications, where pulsatile flows, patient‑specific geometries, and multi‑scale interactions must be captured with precision. As a result, there is a growing need for alternative methodologies that can balance accuracy, efficiency, and scalability. \cite{DataDrivenFluidSimulation} 

Recent advances in machine learning, and particularly in deep learning, have introduced new paradigms for fluid dynamics simulation. \cite{DNN_fluid_dynamics} Among these, physics‑informed neural networks (PINNs) have emerged as a promising framework. \cite{OriginalPINNpaper, Emerging_DL_methods} PINNs embed the governing PDEs directly into the loss function of a neural network, thereby enforcing physical consistency while leveraging the flexibility of deep learning. This approach eliminates the need for mesh generation, exploits automatic differentiation for gradient computation, and offers a scalable solution for high dimensionality problems. In the context of fluid dynamics, PINNs have demonstrated the ability to capture complex flow behaviours, including pulsatile and turbulent regimes, with reduced computational overhead compared to conventional CFD. \cite{MLforCFD}

The integration of PINNs into fluid dynamics research represents a significant step toward bridging data‑driven and physics‑based modelling. By combining the rigour of physical laws with the adaptability of neural networks, PINNs provide a computationally efficient and accurate alternative for simulating flows in challenging scenarios. \cite{PINNFluidsPInit} Their potential impact spans multiple domains, from biomedical engineering, where they can aid in the study of vascular haemodynamics, to industrial applications requiring real‑time flow prediction. This study builds upon these developments by implementing a three‑dimensional PINN framework for the analysis of blood flow in the human aorta, highlighting its advantages over traditional CFD approaches in terms of efficiency, scalability, and physical fidelity. \cite{PINNsForFluids, SUNcomPINNvsVol}

The study of blood flow within the human aorta is of great relevance in the evaluation of Abdominal Aortic Aneurysms (AAAs) and its associated comorbidity and mortality. AAAs are localized dilatations of the abdominal aorta characterized by progressive structural deterioration of the vascular wall, which predisposes the vessel to further enlargement and, ultimately, rupture. \cite{BiomechanicsAAA} The risk of rupture is strongly correlated with aneurysm diameter and growth rate, and rupture events are associated with exceptionally high mortality \cite{AAARuptureRisk}. Epidemiological data indicate that AAAs represent one of the leading causes of death in individuals over 55 years of age in both Europe and the United States of America, ranking between the twelfth and fifteenth most common causes of mortality in this demographic.\cite{AAAGrowthSpain}

Consequently, accurate modelling of aortic haemodynamics is essential for improving risk stratification, supporting clinical decision‑making, and informing therapeutic strategies aimed at reducing aneurysm‑related mortality. In this context, the primary objective of the present work is to develop a model capable of predicting the pressure exerted by the internal blood flow on the wall of the abdominal aorta. To this end, the PINN framework is used to estimate both velocity and pressure fields throughout the domain, including interior points, which are necessary for reconstructing the full haemodynamic flow pattern.

The remainder of this article is structured as follows: Section \ref{sec:Background} provides a review of related work on simpler and similar geometries, Section \ref{sec:Methodology} details the procedure for the present work, including mathematical background, and describing the dataset used, technical implementation of the model and the validation method. Section \ref{sec:Results} presents the results obtained from the trained PINN, supported by visualisations of the predicted haemodynamic fields. Section \ref{sec:Disscusion} discusses these findings in relation to existing literature, examines the validation metrics, and outlines the main limitations of the approach. Finally, Section \ref{sec:Limitations} shows the main limitations of the current model and show potential improvements, and finally Section \ref{sec:Conclusion} summarises the conclusions of the study and identifies potential directions for future research.

\section{Related Works} \label{sec:Background}

This section includes, structured in several subsections, an introduction to the governing fluid‑dynamics equations and their treatment within the model, a review of previous PINN‑based studies on simplified geometries.

\subsection{Previous works on simpler geometries} \label{sec:PW-simpler-geom}

Building on this theoretical foundation, numerous studies have explored how the PINN framework performs when applied to fluid‑dynamic problems in simplified geometries. These lower‑dimensional settings provide an ideal environment for evaluating training strategies, loss‑balancing techniques, and the overall ability of PINNs to reproduce solutions of the Navier–Stokes equations without relying on mesh‑based discretization. Early demonstrations, such as the work of Raissi et al. \cite{OriginalPINNpaper}, showed that PINNs can accurately recover solutions to nonlinear PDEs in one and two dimensions, establishing a baseline for subsequent investigations into more specialized flow configurations.

Several recent contributions have focused specifically on 2D flow scenarios, using them as controlled testbeds to assess the strengths and limitations of the method. Botarelli et al. \cite{PINNsForFluids} applied PINNs to two‑dimensional Navier–Stokes problems in complex but still planar geometries, illustrating how the mesh‑free formulation facilitates the treatment of irregular boundaries and heterogeneous flow conditions. Their results highlight the flexibility of the approach in settings where traditional CFD methods require careful meshing or stabilization. 

Complementing this, Wong et al. \cite{multicasePINN2D} introduced a multi‑case training strategy in which a single PINN is exposed to multiple tube‑flow configurations during training. This approach significantly improves generalization across different 2D geometries and reduces the computational cost associated with training separate models for each case, demonstrating how shared physical structure can be leveraged within the PINN framework.

In the context of haemodynamics, simplified arterial geometries have also served as an important proving ground for PINN‑based modelling. While classical CFD studies, such as Kabir et al. \cite{Num2Dstenosis}, provide detailed simulations of pulsatile blood flow in normal and stenosed arteries, more recent PINN-oriented work has sought to integrate physical constraints and data within a unified optimization process. Liu et al. \cite{LiuPINN1D2D} proposed a variable‑separated PINN architecture equipped with adaptive loss weighting to address the imbalance between PDE residuals and boundary‑condition terms, a common challenge in blood‑flow simulations. Their results demonstrate improved convergence and accuracy in simplified flow domains, underscoring the importance of carefully designed loss‑balancing strategies when applying PINNs to physiological problems.

Taken together, these studies show that simple 1D and 2D geometries play a crucial role in the development and refinement of PINN methodologies for fluid dynamics. By first validating the approach in controlled settings, it is possible to systematically identify issues such as stiffness in the loss landscape, sensitivity to boundary‑condition enforcement, or slow convergence. The insights gained from these lower‑dimensional experiments form the basis for extending PINNs to more demanding three‑dimensional and time‑resolved cardiovascular applications, which will be discussed in the following section.

\subsection{Previous works on 3D and 4D geometries} \label{sec:PW-complex-geom}

The use of physics-informed neural networks for fully three-dimensional and time-resolved (4D) flow problems has grown substantially in recent years, particularly in the context of cardiovascular modelling and medical imaging. 

One of the earliest applications in this direction was presented by Fathi et al. \cite{4DFlowMRI}, who developed a physics-informed deep learning framework for super--resolution and denoising of 4D-Flow MRI data. 

By embedding the Navier-Stokes equations into the reconstruction process, their method enhances noisy or low--resolution velocity fields while preserving physical consistency, demonstrating the potential of PINNs as physics--aware post--processing tools for clinical imaging.

Another important line of research focuses on using PINNs to infer unmeasured haemodynamic quantities from sparse or incomplete 4D flow data. Kissas et al. \cite{MLcardflow4D} showed that PINNs can reconstruct arterial pressure fields from non-invasive 4D-Flow MRI velocity measurements by enforcing the incompressible Navier-Stokes equations as soft constraints during training. Their work illustrates how PINNs can recover clinically relevant but unobservable quantities, such as pressure gradients, without requiring full boundary condition specification or mesh generation, offering an alternative to traditional CFD pipelines.

Beyond data assimilation, several studies have explored PINNs as direct solvers for three-dimensional Navier-Stokes problems. Alzhanov et al. \cite{3DcoronarySim} proposed a fully 3D PINN framework for simulating blood flow in patient-specific coronary artery trees, integrating imaging data with physics-based constraints to estimate fractional flow reserve (FFR). Their results show good agreement with CFD simulations and invasive measurements. 

Similarly, Heger et al. \cite{PINNs3Dvehicles} investigated the use of physics-informed deep learning to predict parametric 3D flow fields from boundary data, demonstrating that PINNs can serve as efficient surrogates for families of flow solutions parametrized by geometry or boundary conditions.

Methodological advances have also contributed to improving the robustness of PINNs in high-dimensional flow settings. Liu et al. \cite{LiuPINN1D2D} introduced a variable-separated PINN architecture with adaptive loss weighting to address the imbalance between PDE residuals and boundary constraints in blood-flow simulations. Their approach improves convergence and accuracy in complex flow regimes, highlighting the importance of architectural and optimization strategies when extending PINNs to 3D and 4D domains. Afterwards, in \cite{LIU2026110011} the same team extends the Integral Conservation Physics‑Informed Neural Networks (ICPINNs) framework to transient blood‑flow simulations in patient‑specific thoracic aortas, incorporating the integral form of the Navier–Stokes equations and Monte Carlo integration to enhance physical fidelity. The authors conduct the first systematic comparison of multiple neural network architectures.

In \cite{PINN_DeepONet_comp} Cruz-Gonzalez et al. present a comparative study of classical PINNs, Deep Operator Networks (DeepONets), and their physics-informed variants (PI-DeepONets) for simulating blood flow in an idealised 3D abdominal aortic aneurysm model. The authors embed the steady Navier–Stokes equations into the learning process and systematically assess accuracy and computational efficiency against high-fidelity CFD benchmarks. While in \cite{PINN4Daorta_comp} Zhang et al. propose a 4D haemodynamic prediction framework that integrates CFD‑generated datasets with deep learning models based on point‑cloud representations and physics‑informed neural networks. The study constructs two large 4D vascular datasets—one including fine coronary branches and another focused on the abdominal aorta—and evaluates multiple architectures, including PointNet, PointNet++, and PINN‑enhanced variants, to determine the optimal framework for different vascular morphologies.

Overall, these studies demonstrate that PINNs have evolved into a versatile tool for 3D and 4D fluid-dynamic modelling, capable of performing simulation, data assimilation, super-resolution, and surrogate modelling within a unified framework. They also emphasize the key challenges, including training stability, loss balancing, and scalability; that motivate ongoing research and inform the design of the present work.

\subsection{PINNs for Abdominal Aortic Aneurysm Modelling} \label{sec:PW-AAA}

In recent years, physics-informed neural networks have increasingly been applied to the study of abdominal aortic aneurysms (AAAs), where accurate haemodynamic modelling is essential for understanding aneurysm progression and rupture risk. Compared to simplified vascular domains, AAA geometries present additional challenges due to their complex morphology, pulsatile flow conditions, and the presence of highly heterogeneous velocity and pressure distributions. These characteristics make AAAs a particularly suitable benchmark for evaluating the capabilities and limitations of PINN-based haemodynamic solvers.

Recent studies have explored the use of PINNs for simulating blood flow in aneurysmal geometries and reconstructing clinically relevant haemodynamic quantities. Cruz-González et al. \cite{CruzGonzalez_2026} presented a comparative framework combining classical PINNs, Deep Operator Networks (DeepONets), and physics-informed DeepONet variants for the simulation of blood flow in idealized three-dimensional abdominal aortic aneurysm geometries. Their work demonstrated that operator-learning approaches can improve computational efficiency while preserving the physical consistency imposed by the Navier-Stokes equations, highlighting the potential of hybrid architectures for vascular-flow modelling.

Beyond purely fluid-dynamic simulations, other works have investigated the interaction between blood flow and the arterial wall developing a coupled fluid-structure interaction (FSI) framework, integrating PINNs with conventional computational fluid dynamics methods to analyse pulsatile flow in arterial aneurysms \cite{Rehman_2025}. Their results showed that incorporating vessel-wall deformation provides a more realistic representation of aneurysmal haemodynamics and may improve the assessment of biomechanical factors associated with aneurysm growth and rupture.

Related efforts have also focused on improving the prediction of complex haemodynamic patterns in patient-specific vascular geometries such as a four-dimensional haemodynamic prediction framework combining CFD-generated datasets with physics-informed neural networks and point-cloud representations \cite{Zhang_2023}. By analysing different vascular morphologies, including abdominal aortic configurations, the study demonstrated that PINN-enhanced models can capture temporally resolved flow dynamics while reducing the computational cost associated with traditional CFD simulations.

Similarly, another investigation shows the combined use of CFD and PINNs for rupture-risk prediction in thoracoabdominal aneurysms through fluid-structure interaction analysis \cite{Chen_2025}. Their findings suggest that physics-informed learning frameworks may contribute to improved biomechanical indicators for aneurysm assessment, particularly when integrated with patient-specific anatomical and flow information.

Despite these advances, several studies report persistent limitations related to training stability, convergence behaviour, sensitivity to hyperparameter selection, and scalability in complex or turbulent flow regimes \cite{Chuang_2022, Aghaee_2025}. Although PINNs reduce the dependence on mesh generation and large simulation datasets, their computational cost during training may remain substantial, and they do not consistently outperform established CFD methods \cite{PINNsForFluids}. These challenges continue to motivate research into improved architectures, adaptive training strategies, and more robust optimization methods for large-scale haemodynamic applications.

\subsection{Contributions} \label{sec:Contributions}

The present study advances the state of research on Physics‑Informed Neural Networks by demonstrating their applicability to fully 3D haemodynamic simulations in the abdominal aorta. While previous work has largely focused on simplified geometries, steady‑state conditions, or idealized flow regimes, and there is not yet a wide consolidated literature on more complex scenarios such as in aortic aneurysms, our model shows that PINNs can successfully reconstruct pressure and velocity fields in anatomically realistic vascular complex domains without relying on mesh‑based discretization. This contributes to the growing evidence that PINNs can serve as a computationally efficient alternative to conventional CFD, particularly in scenarios where boundary conditions are uncertain or clinical data are sparse.

A second key contribution lies in the systematic evaluation of the model’s physical consistency through continuity and momentum residuals, providing quantitative evidence of the reliability of the PINN‑based predictions. By comparing these metrics with those reported in the literature for both simplified and complex vascular geometries, the study positions its results within the broader research landscape and demonstrates that the proposed framework achieves error levels comparable to, or in some cases lower than, existing PINN‑based haemodynamic models. This reinforces the feasibility of using PINNs for clinically relevant flow estimation tasks, even in the absence of a CFD‑derived ground truth.

Finally, the study identifies and articulates several methodological pathways for future development, including the incorporation of non-Newtonian fluid modelling, improved generalisation across diverse vascular morphologies, and the exploration of more advanced PINN architectures such as CNN‑PINNs or GCN‑PINNs. By outlining these directions, the work not only situates itself within the current trajectory of PINN research but also provides a roadmap for enhancing the accuracy, robustness, and clinical applicability of physics‑informed deep learning in cardiovascular biomechanics.

\section{Methodology} \label{sec:Methodology}

The current section contains an extensive explanation of the theoretical background and the technical implementation needed for the current work. The section begins with mathematical explanation in \ref{sec:Math-background}, where the core concepts of fluid dynamics have been analysed, followed in Sec. \ref{sec:Dataset} by a review of the original dataset used for the present work, then in Sec. \ref{sec:PINN-imp} the technical architecture and implementations of the PINN are reviewed, and lastly in Sec. \ref{sec:Validation} the model validation method is presented.

\subsection{Mathematical background} \label{sec:Math-background}

Beginning with the mathematical background needed to understand the physical principles and constraints of fluid dynamics, this sections has been divided into two separate subsections. First, the core concepts and deduction of the Navier-Stokes equations shall be reviewed in Subsect. \ref{sec:Bg-intro}, continuing in Subsect. \ref{sec:4D-ecuations} with the expansion of the 3D time dependent Navier-Stokes equations and deduction of the residual therms that will be used in the PINN composed loss function.

\subsubsection{Introducing fluid dynamics into PINNs} \label{sec:Bg-intro}

The foundation of fluid dynamics lies in the system of PDEs that explain the motion of viscous fluids, namely the Navier-Stokes equations \cite{Navier_Stokes_analysis}. These equations, derived from the principles of conservation of mass and momentum and provide the mathematical framework for describing a wide range of flow phenomena, from laminar to turbulent regimes. As such, they constitute the cornerstone of both theoretical analysis and computational modelling in fluid mechanics.

The Navier–Stokes equations can be derived as a particular case of the Cauchy momentum equation. By expressing the Cauchy stress tensor as the sum of an isotropic pressure term and a viscous stress term, the governing equations reduce to the convective form of the momentum balance for a Newtonian fluid. This formulation provides the foundation from which the Navier-Stokes equations are obtained, linking the general principles of continuum mechanics to the specific description of viscous fluid motion.

\begin{equation}
\label{CauchyMom}
\rho \frac{ \text{D} \boldsymbol{u}}{\text{D}t} = - \nabla p + \nabla \cdot \boldsymbol{\tau} + \rho \boldsymbol{a} ,
\end{equation}

where $\tfrac{\text{D}}{\text{D}t}$ is the material derivative, defined as $\tfrac{\partial}{\partial t} + \boldsymbol{u} \cdot \nabla$, $\rho$ is the fluid density, and $p$ is the pressure, and $\boldsymbol{a}$ represents all external acceleration acting over the fluid. 

From this point onward, several assumptions are introduced to derive the specific form of the Navier–Stokes equations required for this problem: 

\begin{enumerate} 
\item Blood is treated as a Newtonian fluid (constant viscosity). 
\item The fluid density is considered constant (incompressible). 
\item The flow takes into account voticity and turbulence due to blood's typical Reynolds number. \end{enumerate}

Under these assumptions, the Navier–Stokes equations for an incompressible laminar flow can be derived from the Cauchy momentum equation (\ref{CauchyMom}), yielding the conservation of momentum and the conservation of mass \cite{PINNlaminarFlow}:

The momentum conservation equation for each component can be expressed as follows:
    \begin{equation}
    \label{GenMomentum}
        \frac{\partial u_i}{\partial t} + \sum_ju_j\frac{\partial u_i}{\partial x_j} = - \frac{1}{\rho}\frac{\partial p}{\partial x_i}  + \frac{1}{\rho} \sum_i\frac{\partial \tau_{ij}}{\partial x_j},
    \end{equation}

where $u_i$ is the $i^{th}$ component of the velocity vector, $x_i$ is the $i^{th}$ component of the coordinate system. Lastly, $\tau_{ij}$ represents the viscous stress tensor, whose expression for an incompressible Newtonian fluid is:

\begin{equation}
\label{ViscStressT}
    \tau_{ij} = \mu \left( \frac{\partial u_i}{\partial x_j} + \frac{\partial u_j}{\partial x_i}   \right) \quad\Rightarrow\quad\sum_j\frac{\partial \tau_{ij}}{\partial x_j}=\mu\sum_j\frac{\partial^2u_i}{\partial x^2_j}.
\end{equation}

Meanwhile, the mass conservation or continuity equation is more simply expressed as:
    \begin{equation}
    \label{ContEq}
        \nabla\cdot \boldsymbol{u} \quad\Rightarrow\quad\frac {\partial u_i}{\partial x_i} = 0.
    \end{equation}

Taking both eq. \ref{GenMomentum} and eq. \ref{ContEq} as the governing equations for a particle system in an incompressible Newtonian laminar fluid, the Physics‑Informed Neural Network framework incorporates these equations directly into the loss function. The loss terms are formulated so that each governing equation contributes explicitly to the optimization problem solved during backpropagation, where automatic differentiation is used to compute the required derivatives \cite{PINNconcept}.

A PINN augments the traditional data‑driven loss function with additional terms that enforce the governing physical laws. These terms are evaluated at a set of collocation points distributed throughout the spatiotemporal domain, where the neural network predictions are required to satisfy the momentum and continuity equations. Each residual corresponding to the deviation from the exact PDE solution is incorporated into the loss function, ensuring that the optimization process penalizes physically inconsistent predictions \cite{OriginalPINNpaper}.

During backpropagation, automatic differentiation is used to compute all spatial and temporal derivatives appearing in the PDE residuals. This eliminates the need for numerical discretization and allows the network to learn a solution that is continuously differentiable across the domain. As a result, the optimization algorithm simultaneously minimizes the data mismatch and the physics‑based residuals, guiding the network toward solutions that adhere to both the observed measurements and the underlying fluid dynamics. This unified treatment of data and physics is one of the key advantages of PINNs, enabling them to generalize well even in regimes where training data are sparse or noisy.

To complete the physical constraints, the boundary conditions of the geometry must also be incorporated into the loss function, ensuring that the neural network satisfies not only the governing equations within the domain but also the prescribed behaviour at the inlet, outlet, and vessel walls. These terms enforce conditions such as imposed velocity profiles, pressure values, or no‑slip constraints, depending on the specific configuration of the problem. By embedding both the interior physics and the boundary information into a unified optimization framework, the PINN is guided toward solutions that remain fully consistent with the physical requirements of the system.

Therefore, the final loss function is composed as follows:

\begin{equation}
\label{StationaryLossFunc}
    L = L_{\text{m}} + L_{\text{c}} + L_{\text{bc}} + L_{\text{ic}},
\end{equation}

where $L_\text{m}$ is the momentum equation, the $L_\text{c}$ corresponds to the continuity equation, and the $L_\text{bc}$ and $ L_{\text{ic}}$ add the boundary and initial conditions. All terms should be properly arranged so that they converge to zero individually.

\subsubsection{Developing 4D fluid equations} \label{sec:4D-ecuations}

For a 3D-spatial non static case of an incompressible flow, the momentum equation (\ref{GenMomentum}) can be expanded as follows:

\begin{equation}
\label{3DMomentumEq}
\left\lbrace
\begin{array}{ll}
\frac{\partial u_x}{\partial t} + u_x \frac{\partial u_x}{\partial x} + u_y \frac{\partial u_x}{\partial y} + u_z \frac{\partial u_x}{\partial z} = - \frac{1}{\rho} \frac{\partial p}{\partial x} + \frac{\mu}{\rho} \left( \frac{\partial^2 u_x}{\partial x^2} + \frac{\partial^2 u_x}{\partial y^2} + \frac{\partial^2 u_x}{\partial z^2} \right), \\
\frac{\partial u_y}{\partial t} + u_x \frac{\partial u_y}{\partial x} + u_y \frac{\partial u_y}{\partial y} + u_z \frac{\partial u_y}{\partial z} = - \frac{1}{\rho} \frac{\partial p}{\partial y} + \frac{\mu}{\rho} \left( \frac{\partial^2 u_y}{\partial x^2} + \frac{\partial^2 u_y}{\partial y^2} + \frac{\partial^2 u_y}{\partial z^2} \right), \\
\frac{\partial u_z}{\partial t} + u_x \frac{\partial u_z}{\partial x} + u_y \frac{\partial u_z}{\partial y} + u_z \frac{\partial u_z}{\partial z} = - \frac{1}{\rho} \frac{\partial p}{\partial z} + \frac{\mu}{\rho} \left( \frac{\partial^2 u_z}{\partial x^2} + \frac{\partial^2 u_z}{\partial y^2} + \frac{\partial^2 u_z}{\partial z^2} \right),
\end{array}
\right.
\end{equation}

and the continuity equation can be expanded as:

\begin{equation} 
\label{3DContEq}
\frac{\partial u_x}{\partial x} + \frac{\partial u_y}{\partial y} + \frac{\partial u_z}{\partial z} = 0.
\end{equation}
    
After expanding the Navier–Stokes equations in three spatial dimensions, it is convenient to rewrite the system in compact vector form. Let 
$\mathbf{u}(x,y,z,t) = (u_x,u_y,u_z)$ denote the velocity field and 
\(p(x,y,z,t)\) the pressure field, defined on a spatial domain 
$\Omega \subset \mathbb{R}^3$ and a time interval $t \in [0,T]$. 
For an incompressible Newtonian fluid with constant density $\rho$ and 
dynamic viscosity $\mu$, the governing equations can be written as:

\begin{equation}
\label{NSVectorForm}
\frac{\partial \mathbf{u}}{\partial t} 
+ (\mathbf{u} \cdot \nabla)\mathbf{u} 
= -\frac{1}{\rho}\nabla p 
+ \nu \nabla^2 \mathbf{u},
\qquad
\nabla \cdot \mathbf{u} = 0,
\end{equation}

where \(\nu = \mu / \rho\) is the kinematic viscosity.

In the four–dimensional spatio–temporal setting considered here, the PINN takes as input the coordinates $(x,y,z,t)$ and outputs the corresponding flow variables:
\begin{equation}
(x,y,z,t) \;\mapsto\; 
\big(u_x(x,y,z,t),\, u_y(x,y,z,t),\, u_z(x,y,z,t),\, p(x,y,z,t)\big).
\end{equation}

From these outputs, the residuals of the momentum and continuity equations are constructed at each collocation point. For the three momentum components, the residuals are defined as:

\begin{equation}
\label{MomResiduals}
\left\lbrace
\begin{aligned}
R_{m,x} &= 
\frac{\partial u_x}{\partial t} 
+ u_x\frac{\partial u_x}{\partial x} 
+ u_y\frac{\partial u_x}{\partial y} 
+ u_z\frac{\partial u_x}{\partial z} 
+ \frac{1}{\rho}\frac{\partial p}{\partial x} 
- \nu \nabla^2 u_x, \\
R_{m,y} &= 
\frac{\partial u_y}{\partial t} 
+ u_x\frac{\partial u_y}{\partial x} 
+ u_y\frac{\partial u_y}{\partial y} 
+ u_z\frac{\partial u_y}{\partial z} 
+ \frac{1}{\rho}\frac{\partial p}{\partial y} 
- \nu \nabla^2 u_y, \\
R_{m,z} &= 
\frac{\partial u_z}{\partial t} 
+ u_x\frac{\partial u_z}{\partial x} 
+ u_y\frac{\partial u_z}{\partial y} 
+ u_z\frac{\partial u_z}{\partial z} 
+ \frac{1}{\rho}\frac{\partial p}{\partial z} 
- \nu \nabla^2 u_z,
\end{aligned}
\right.
\end{equation}

and the residual of the continuity equation enforcing the incompressibility of the model is expressed as:

\begin{equation}
\label{ContResidual}
R_c = 
\frac{\partial u_x}{\partial x} 
+ \frac{\partial u_y}{\partial y} 
+ \frac{\partial u_z}{\partial z}.
\end{equation}

All the previously expanded residuals are evaluated at a set of interior collocation points 
$\{(x_f^i,y_f^i,z_f^i,t_f^i)\}_{i=1}^{N_f}$ where the corresponding physics–based loss terms are defined as: 
\begin{equation}
\label{LmLc}
L_{\text{m}} = \frac{1}{N_f}\sum_{i=1}^{N_f} 
\left( R_{m,x}^{(i)2} + R_{m,y}^{(i)2} + R_{m,z}^{(i)2} \right),
\qquad
L_{\text{c}} = \frac{1}{N_f}\sum_{i=1}^{N_f} R_{c}^{(i)2}.
\end{equation}

To close the problem, suitable initial and boundary conditions are imposed. 
At the inlet, a pulsatile pressure condition is prescribed as:
\begin{equation}
\label{InletPressure}
P_{\text{in}}(t) = 
P_{\min} + P_{\Delta} \sin^2\left( 2\pi f' t \right),
\end{equation}

where $P_\Delta = P_{\max} - P_{\min}$ (so that it is a continuous, smooth function and in the range of $[P_{\min}, P_{\max}]$), $f'=f/2$ and $f$ is the frequency of the pulse. On the other hand, a static pressure $P_{\text{min}}$ is enforced at the outlet. 
The inlet velocity profile is assumed parabolic, pulsatile and always greater than $0$,

\begin{equation}
\label{InletVelocity}
u_z(r,t) = 
v_{\text{ref}}\cdot 
\left[ 1 - \left( \frac{r}{R_{\max}} \right)^2 \right]\cdot\left[(1-v_{\text{min}})\sin^2\left( 2\pi f' t \right)+v_{\text{min}}\right],
\end{equation}

where $r$ is the radial distance from the vessel centerline. 
A no–slip condition is applied on the vessel walls,

\begin{equation}
\mathbf{u} = \mathbf{0} 
\quad \text{on } \partial\Omega_{\text{wall}}.
\end{equation}

The initial conditions specify the state of the system at $t = 0$. The inlet pressure is initialized at its minimum value, (diastolic phase) for a soft start,

\begin{equation}
P(x,y,z,0) = P_{\min} 
\quad \text{on } \Gamma_{\text{in}},
\end{equation}

and the velocity field is initialized as:
\begin{equation}
u_x(x,y,z,0) = 0, \qquad
u_y(x,y,z,0) = 0, \qquad
u_z(x,y,z,0) = v_{\text{min}}(x,y,z).
\end{equation}

The boundary and initial conditions contribute with additional loss terms 
$L_{\text{bc}}$ and $L_{\text{ic}}$, defined as mean–squared errors between the network 
predictions and the prescribed values. Extending the stationary loss 
definition in \eqref{StationaryLossFunc} to the time–dependent 
three–dimensional case, the total loss becomes:

\begin{equation}
\label{4DLoss}
L = L_{\text{m}} + L_{\text{c}} + L_{\text{bc}} + L_{\text{ic}},
\end{equation}

with each contribution arranged to converge to zero during training, ensuring 
that the PINN satisfies the Navier–Stokes equations, incompressibility, and 
all imposed physical constraints throughout the full four–dimensional domain.

\subsection{Dataset} \label{sec:Dataset}

The original data come from computed axial tomography scans processed by the biomedical engineering team at Avamed Synergy, who, as collaborators of the ENIA Research Chair of Artificial Intelligence at the University of Alicante, have provided their datasets for the development of the present PINN model. The biomedical engineering team extracts a 3D object containing the geometry of the abdominal aorta, which is then stored in STL file format.

\begin{figure} [H]
    \centering
    \includegraphics[width=0.4\linewidth]{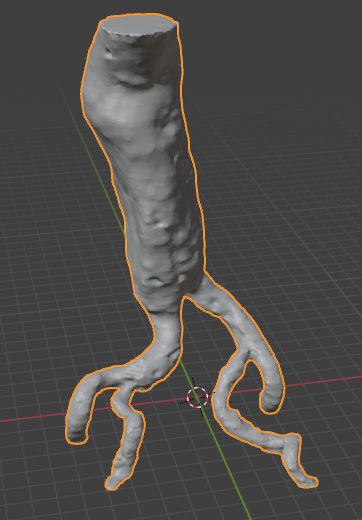}
    \caption{Original STL view in blender}
    \label{fig:blender-STL}
\end{figure}

Given that the model relies exclusively on point clouds as input, we extract only the positional data, disregarding all additional mesh information such as connectivity vectors between nodes. This allows us to isolate the points that define the vessel surface, that is, the locations where the prescribed boundary conditions will be applied.

In addition, an automatic detection system for identifying inlet and outlet regions of the flow was implemented, since it is essential to distinguish the areas where specific boundary conditions must be imposed for blood‑flow entry and exit. These include the direction, profile, and magnitude of the inlet velocity, constraints on lateral velocity components, and inlet and outlet pressure conditions.

This automatic detection system is based on predefined numerical rules, taking into account that, for the region of interest considered, the inlet and outlet zones consistently lie along the boundaries of the domain. The upper boundary is treated as the inlet region, whereas the lateral and lower boundaries are considered potential outlets. Minimum distance thresholds and additional constraints were also introduced to prevent partial contacts with the domain boundary from being erroneously classified as outlet regions.

\begin{figure} [H]
    \centering
    \begin{subfigure}[b]{0.45\linewidth}
        \includegraphics[width=\linewidth]{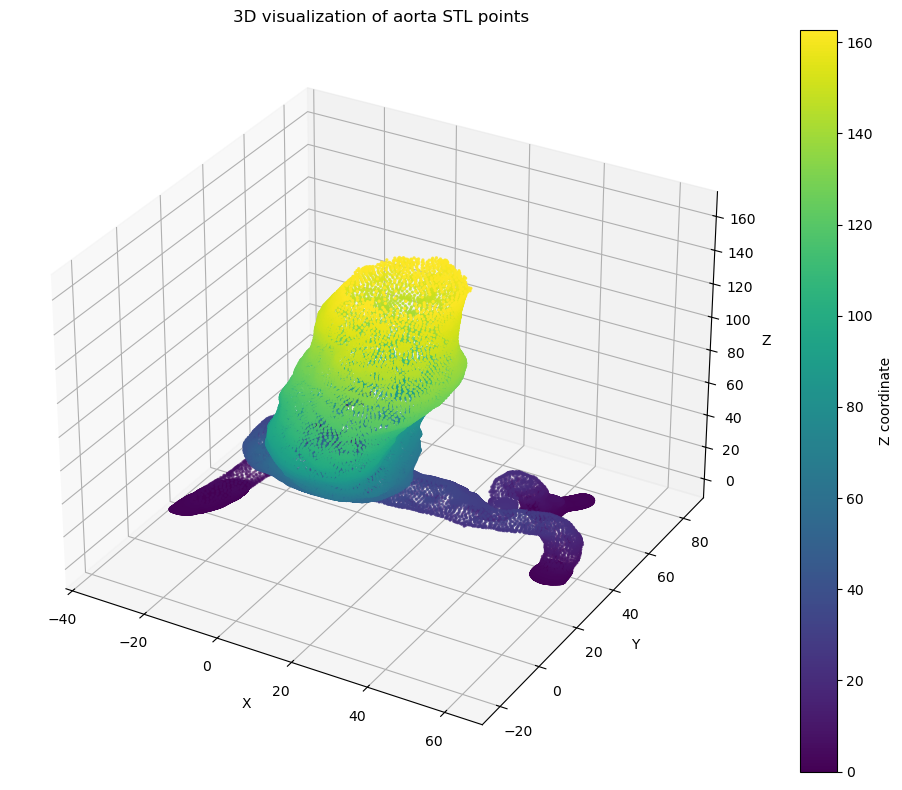}
    \end{subfigure}
    \begin{subfigure}[b]{0.4\linewidth}
        \includegraphics[width=\linewidth]{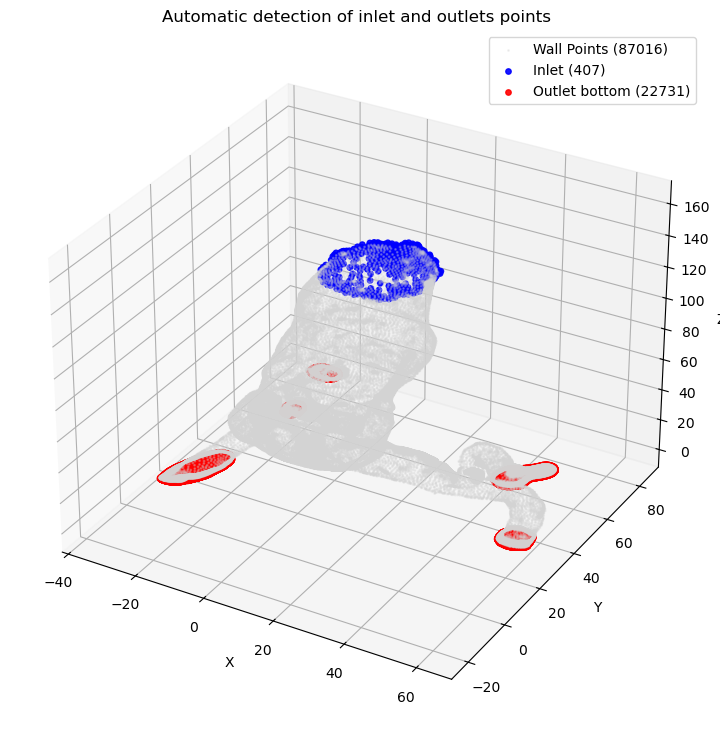}
    \end{subfigure}
    \caption{Points extracted from STL and automatic inlet and outlet detection}
    \label{fig:auto-inlet-outlet}
\end{figure}

\subsection{PINN implementation} \label{sec:PINN-imp}
The PINN developed in this work has been implemented using the Python programming language, specifically through the specialized DeepXDE library built on a PyTorch backend. Following the examination of the governing physical equations presented in the previous sections, we now turn to the concrete implementation of the practical case under study.
        
DeepXDE is a Python library specifically designed to make physics‑informed neural networks (PINNs) practical, accessible, and flexible for solving a wide range of differential equations. Introduced by Lu et al. (2021) \cite{DeepXDE}, it provides a unified framework in which users can formulate forward and inverse problems directly from their mathematical definitions, while the library handles automatic differentiation, loss construction, and training. 

This library supports ordinary, partial, and even stochastic differential equations, and it accommodates complex geometries through constructive solid geometry. Its design emphasizes compact, mathematically intuitive code, and it incorporates advanced features such as residual‑based adaptive refinement to improve training efficiency. Overall, DeepXDE provides a robust and user‑friendly environment for applying PINNs to real‑world scientific machine‑learning problems.

However, the handling of complex geometries is not fully addressed by the library, making it necessary to develop a set of additional components capable of operating on the available mesh data, converting them into point clouds, and enabling their use within DeepXDE without compromising accuracy.

\begin{figure} [H]
    \centering
    \includegraphics[width=0.8\linewidth]{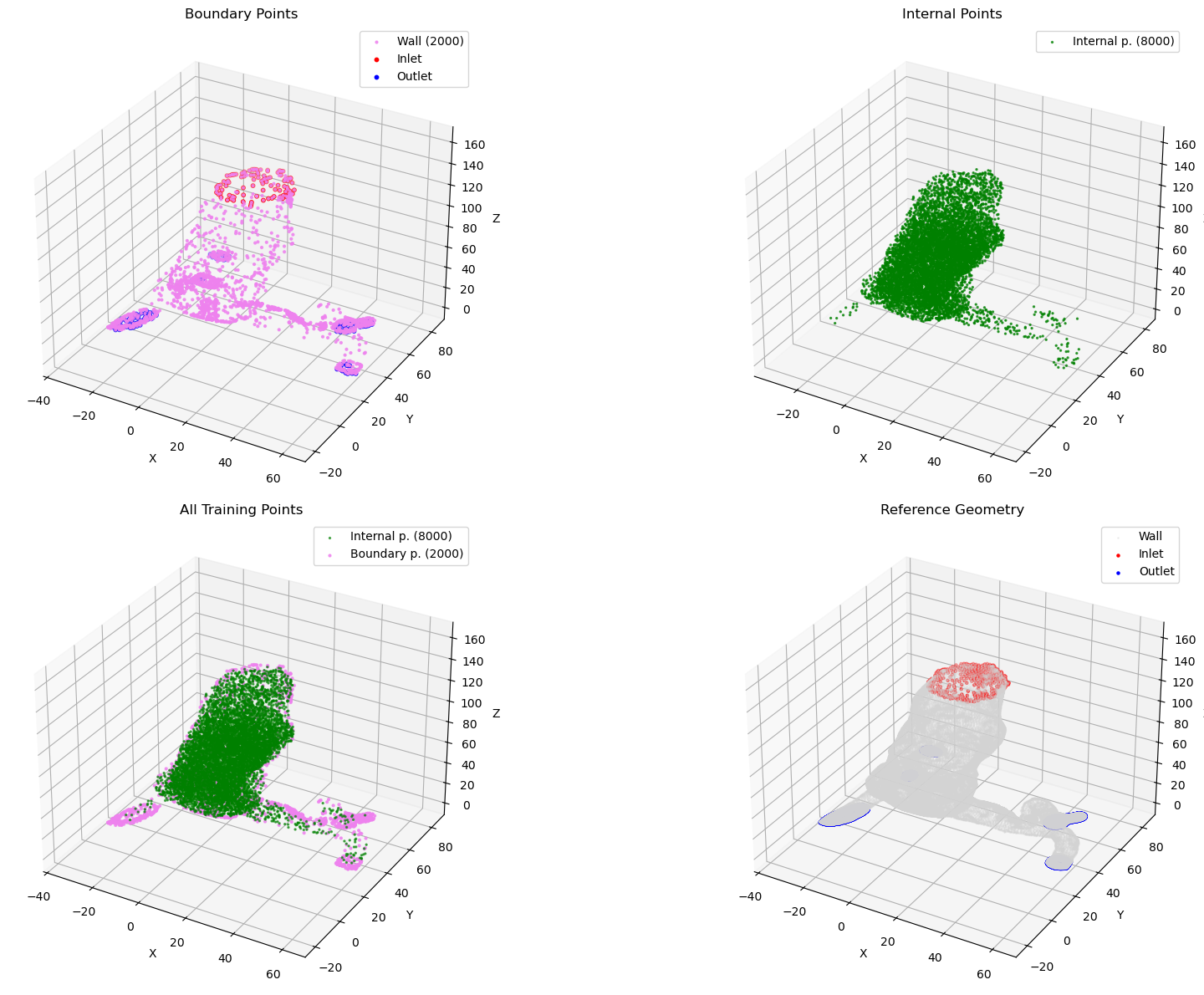}
    \caption{Representation of the inner and boundary points generated to train the model within DeepXDE}
    \label{fig:geometry-points}
\end{figure}

As explained in the previous section, the initial data contain only the points corresponding to the aortic lumen, without including any interior points. For this reason, a key step was the development of a geometry class compatible with DeepXDE that could generate interior points while adapting to a complex and irregular vascular shape. Additionally, it was taken into account that the number of boundary points in the STL file could overload the model during training. To address this question, the custom geometry class performs a random selection of boundary points, while always ensuring a predefined proportion of points belonging to the inlet and outlet regions.

For the generation of interior points, random point clouds are created at different heights, and a verification step is performed using DBSCAN \cite{schubert2017dbscan} to determine whether each point lies inside or outside the region delimited by the boundary surface. Points classified as exterior are discarded. If the number of validated interior points is lower than required, the process is repeated until the desired quantity is reached. Figure \ref{fig:geometry-points} shows an example of the points generated by the specific class designed in order to properly convert the mesh data to data points needed by DeepXDE. 

The fluid‑dynamics equations, together with the initial and boundary conditions, were implemented in the loss function following the procedures established by the library. For the physical parameters of blood, standard average values commonly reported in the specialized literature were adopted: density of 1060 kg/m³, dynamic viscosity of 0.004 $\text{Pa} \cdot \text{s}$, characteristic aortic diameter of 0.02 m, and a maximum inlet velocity of 1 m/s. Maximum inlet pressure is set at mean systolic pressure in adults, 120 mmHg or 15998 Pa, and outlet pressure is set at mean diastolic pressure of 80 mmHg or 10665 Pa \cite{AortaBloodVelocity}.

\begin{figure} [H]
    \centering
    \includegraphics[width=1\linewidth]{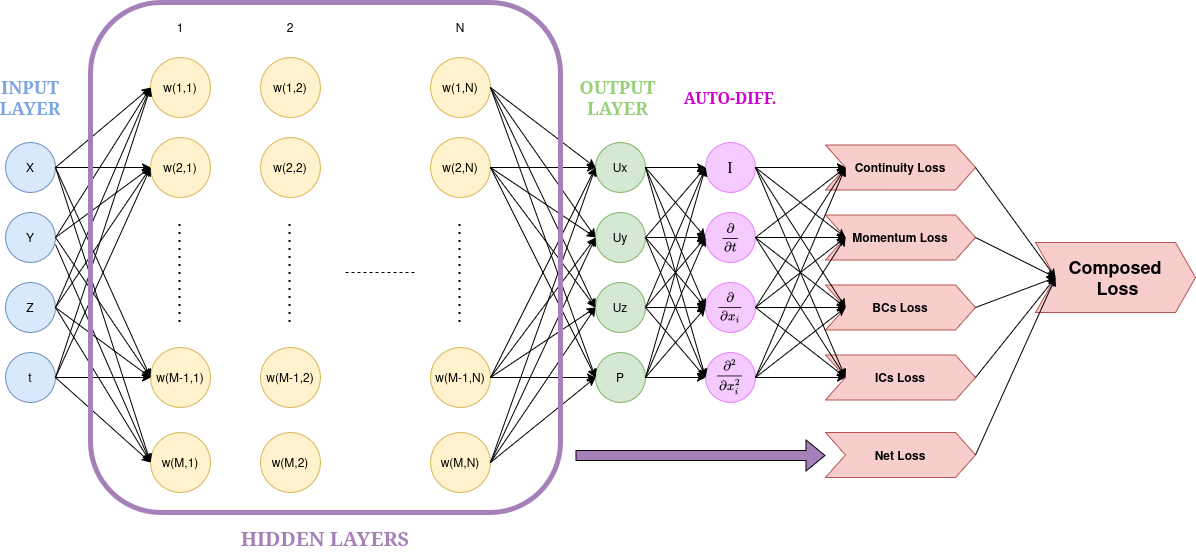}
    \caption{Architecture of a (M,N) net used in a PINN and the composed loss.}
    \label{fig:PINN-Arch}
\end{figure}
        
The feed‑forward neural network (FNN) is structured as a single block of 10 layers with 256 neurons each, according to the architecture shown in Figure \ref{fig:PINN-Arch}, following configurations commonly reported in the related literature \cite{3DcoronarySim}. Training was performed using a definition of 8000 boundary points, 5,000 initial domain points, 2000 boundary points and 1000 test points, granting 10\% of the domain points on the outlets and 5\% on the inlet, together with a temporal simulation window of 180 seconds, using a Hammersley time distribution.

\subsubsection{Problems of the PINN architecture} \label{sec:ModulatedFourierNet}

Standard neural networks are known to suffer from spectral bias, tending to learn low-frequency components faster than high-frequency structures. To mitigate this limitation, Fourier Features based on the work of Tancik et al.~\cite{Tancik_2020_NeurIPS} were incorporated into the input representation. The spatial coordinates are projected into a sinusoidal embedding space:

\begin{equation}
\phi(\mathbf{x}) =
\left[
\sin(2\pi \mathbf{x}\mathbf{B}),
\cos(2\pi \mathbf{x}\mathbf{B})
\right],
\end{equation}

where $\mathbf{B}$ contains randomly initialized frequency matrices with multiple scaling factors. This encoding enables the network to simultaneously capture low, intermediate, and high frequency flow structures.

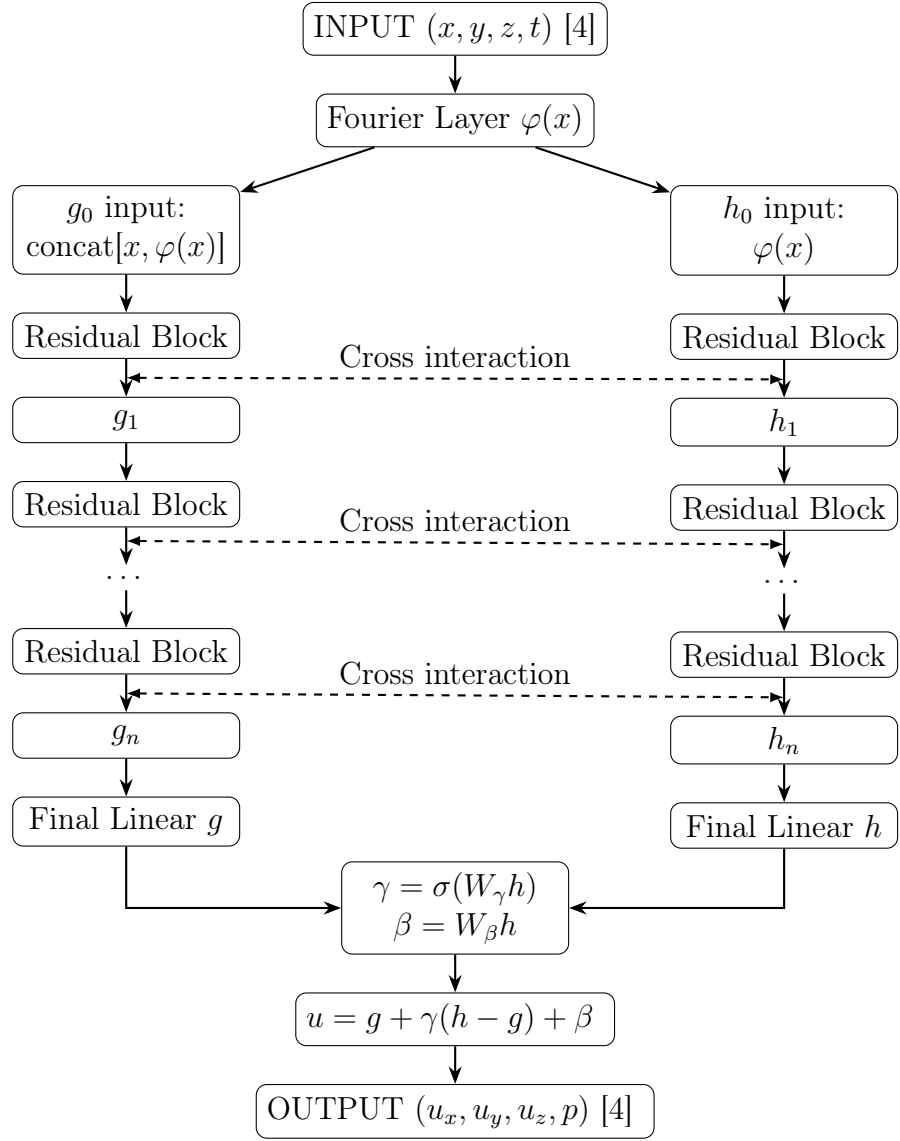
\begin{figure}
\centering
\begin{tikzpicture}[
    node distance=0.5cm,
    box/.style={
        draw,
        rounded corners,
        minimum width=3cm,
        minimum height=0.6cm,
        align=center
    },
    arrow/.style={
        -{Stealth},
        thick
    }
]

\node[box] (input) {INPUT $(x,y,z,t)$ [4]};
\node[box, below=of input] (fourier) {Fourier Layer $\varphi(x)$};
\node[box, below left=0.5cm and of fourier] (g0) {$g_0$ input: \\ concat[$x,\varphi(x)$]};
\node[box, below right=0.5cm and of fourier] (h0) {$h_0$ input: \\ $\varphi(x)$};

\node[box, below=of g0] (gresblock1) {Residual Block};
\node[box, below=of h0] (hresblock1) {Residual Block};

\node[box, below=of gresblock1] (g1) {$g_1$};
\node[box, below=of hresblock1] (h1) {$h_1$};

\node[box, below=of g1] (gresblock2) {Residual Block};
\node[box, below=of h1] (hresblock2) {Residual Block};

\node[below=of gresblock2] (gvdots) {$\dots$};
\node[below=of hresblock2] (hvdots) {$\dots$};

\node[box, below=of gvdots] (gresblockn) {Residual Block};
\node[box, below=of hvdots] (hresblockn) {Residual Block};

\node[box, below=of gresblockn] (gn) {$g_n$};
\node[box, below=of hresblockn] (hn) {$h_n$};

\draw[arrow] (input) -- (fourier);
\draw[arrow] (fourier) -- (g0);
\draw[arrow] (fourier) -- (h0);

\draw[arrow] (g0) -- (gresblock1);
\draw[arrow] (h0) -- (hresblock1);

\draw[arrow] (gresblock1) -- (g1);
\draw[arrow] (hresblock1) -- (h1);

\draw[arrow] (g1) -- (gresblock2);
\draw[arrow] (h1) -- (hresblock2);

\draw[arrow] (gresblock2) -- (gvdots);
\draw[arrow] (hresblock2) -- (hvdots);

\draw[arrow] (gvdots) -- (gresblockn);
\draw[arrow] (hvdots) -- (hresblockn);

\draw[arrow] (gresblockn) -- (gn);
\draw[arrow] (hresblockn) -- (hn);

\coordinate (g1mid) at ($(gresblock1)!0.5!(g1)$);
\coordinate (h1mid) at ($(hresblock1)!0.5!(h1)$);

\draw[latex-latex, dashed, thick] (g1mid) -- node[above] {Cross interaction} (h1mid);

\coordinate (g2mid) at ($(gresblock2)!0.5!(gvdots)$);
\coordinate (h2mid) at ($(hresblock2)!0.5!(hvdots)$);
\draw[latex-latex, dashed, thick] (g2mid) -- node[above] {Cross interaction} (h2mid);

\coordinate (gnmid) at ($(gresblockn)!0.5!(gn)$);
\coordinate (hnmid) at ($(hresblockn)!0.5!(hn)$);

\draw[latex-latex, dashed, thick] (gnmid) -- node[above] {Cross interaction} (hnmid);

\node[box, below=of gn] (gfinal) {Final Linear $g$};
\node[box, below=of hn] (hfinal) {Final Linear $h$};

\node[box, below=of $(gfinal)!0.5!(hfinal)$] (mod) {
$\gamma = \sigma(W_\gamma h)$ \\
$\beta = W_\beta h$
};

\node[box, below=of mod] (u) {
$u = g + \gamma (h-g) + \beta$
};

\node[box, below=of u] (output) {
OUTPUT $(u_x,u_y,u_z,p)$ [4]
};

\draw[arrow] (gn) -- (gfinal);
\draw[arrow] (hn) -- (hfinal);

\draw[arrow] (gfinal) |- (mod);
\draw[arrow] (hfinal) |- (mod);

\draw[arrow] (mod) -- (u);
\draw[arrow] (u) -- (output);

\end{tikzpicture}
\caption{Architecture of the new network (Modulated Fourier Network or MFN).}\label{esq_MFN}
\end{figure}

To improve the ability of the PINN framework to approximate complex solutions of the Navier-Stokes equations in haemodynamic simulations, a custom neural-network architecture was developed combining Fourier Features, parallel specialized branches, residual blocks, and cross-branch interactions (Figure~\ref{esq_MFN}).

The main motivation behind this design is to efficiently represent both smooth global flow structures and localized high-frequency phenomena such as sharp gradients, recirculation regions, and transient pulsatile dynamics.

\subsubsection{Dual-Branch Architecture}

The final architecture consists of two parallel branches:

\begin{itemize}
    \item A \textbf{physics-enriched branch}, receiving the concatenation of physical coordinates and Fourier Features:
\begin{equation}
\mathbf{g}_0 = [\mathbf{x}, \phi(\mathbf{x})],
\end{equation}

which focuses on preserving the global physical structure of the solution.

    \item A \textbf{spectral branch}, operating exclusively on Fourier Features:
\begin{equation}
\mathbf{h}_0 = \phi(\mathbf{x}),
\end{equation}

designed to capture localized high-frequency corrections.
\end{itemize}

\subsubsection{Residual Learning and Cross-Branch Interaction}

Both branches are composed of residual blocks of the form:

\begin{equation}
\mathbf{y} = \mathbf{x} + cF(\mathbf{x}),
\end{equation}

where $F(\mathbf{x})$ is a small internal neural network and $c=0.5$ is a damping factor introduced to improve numerical stability. Residual learning facilitates gradient propagation, stabilizes optimization, and enables progressive refinement of the solution \cite{He_2016_CVPR}.

Unlike standard multi-branch architectures, both branches exchange information after each residual block:

\begin{align}
\mathbf{g}_{k+1} &= \mathbf{g}_k + 0.1\,\mathbf{P}_{h\rightarrow g}(\mathbf{h}_k), \\
\mathbf{h}_{k+1} &= \mathbf{h}_k + 0.1\,\mathbf{P}_{g\rightarrow h}(\mathbf{g}_k),
\end{align}

where $\mathbf{P}_{h\rightarrow g}$ and $\mathbf{P}_{g\rightarrow h}$ are trainable linear projections. This interaction allows the physical branch to incorporate local spectral corrections while maintaining global consistency.

\subsubsection{Adaptive Output Modulation}

The final prediction combines both branches through adaptive modulation coefficients computed from the spectral representation:

\begin{align}
\boldsymbol{\gamma} &= \sigma(\mathbf{W}_{\gamma}\mathbf{h} + \mathbf{b}_{\gamma}), \\
\boldsymbol{\beta}  &= \mathbf{W}_{\beta}\mathbf{h} + \mathbf{b}_{\beta},
\end{align}

leading to the final output:

\begin{equation}
\mathbf{u}
=
(1-\boldsymbol{\gamma})\odot \mathbf{g}
+
\boldsymbol{\gamma}\odot \mathbf{h}
+
\boldsymbol{\beta}.
\end{equation}

This formulation enables the network to adaptively determine the contribution of each branch at every spatial location.

Overall, the proposed architecture combines spectral embeddings, residual refinement, and adaptive multi-scale representations to improve convergence, training stability, and the reconstruction of complex haemodynamic flow patterns in three-dimensional vascular geometries.

\subsection{Model validation} \label{sec:Validation}

The strategy followed in order to validate the model obtained for the blood flow inside the abdominal aorta consists in calculating the residuals for both, continuity and momentum equations. As these equations represent conservation laws for both mass and momentum, the ideal residuals should be nullified, meaning that the Navier-Stokes equations \ref{3DMomentumEq} and \ref{3DContEq} are perfectly fulfilled.

The residuals for those equations were also previously defined in equations \ref{MomResiduals} and \ref{ContResidual}, as they are introduced in the loss function. After the training, the same expressions are applied to the test domain points to verify how close the model is to the best possible outcome.

As previously stated, from the aforementioned residual equations both physics-based loss terms are defined as equation \ref{LmLc} shows, and consequently common error metrics like squared loss function and infinite or maximum loss function can be defined as well, knowing that in the case of the residuals the expected value is 0. Such error metrics are defined as:

\begin{equation}\label{L2 and Linf}
    \begin{split}
    &||R_c||_{L^2} = \sqrt{L_c} \left( \frac{D_{ref}}{u_{ref}} \right)  \qquad ||R_m||_{L^2} = \sqrt{L_m} \left( \frac{D_{ref}}{\rho u^2_{ref}} \right) \\
    & ||R_c||_{L_\infty} =  \max_{j=1,..,N} |R_c^j| \qquad \quad ||R_m||_{L_\infty} =  \max_{j=1,..,N} |\textbf{R}_m^j| 
    \end{split}
\end{equation}

being $L_m$ and $L_c$ the loss terms for momentum and continuity functions respectively, $\textbf{R}_m^j$ and $R_c^j$ are the momentum vector and continuity residuals  corresponding to the $j^{th}$ domain test point. $D_{ref}$ refers to the characteristic diameter of the aorta, and $u_{ref}$ the reference velocity, corresponding to the inlet velocity.

Regarding the physical interpretation of these error metrics, the $L^2$ norm allows us to determine whether there are substantial variations in the average residual error, thereby providing a general measure of the sensitivity in the computation of first and second‑order derivatives. In contrast, the $L_{\infty}$ norm highlights the presence of isolated points where the enforcement of the Navier–Stokes equations may be failing.

\section{Results} \label{sec:Results}

The following section presents the results obtained from the implementation and training of the developed PINN, applying the fluid‑dynamics equations, physical constraints, and network architecture described in the preceding sections. The analysis focuses on the model’s ability to reproduce velocity and pressure fields consistent with the underlying physics, as well as its behaviour during the optimization process, including the convergence of the different loss components and the stability of training. 

Qualitative and quantitative comparisons are also provided to assess the fidelity of the model with respect to the expected characteristics of a 3D pulsatile blood‑flow regime. Taken together, these results allow us to evaluate the feasibility and limitations of the proposed PINN‑based approach for haemodynamic simulation.

\begin{figure}[H]
    \includegraphics[width=0.6\linewidth]{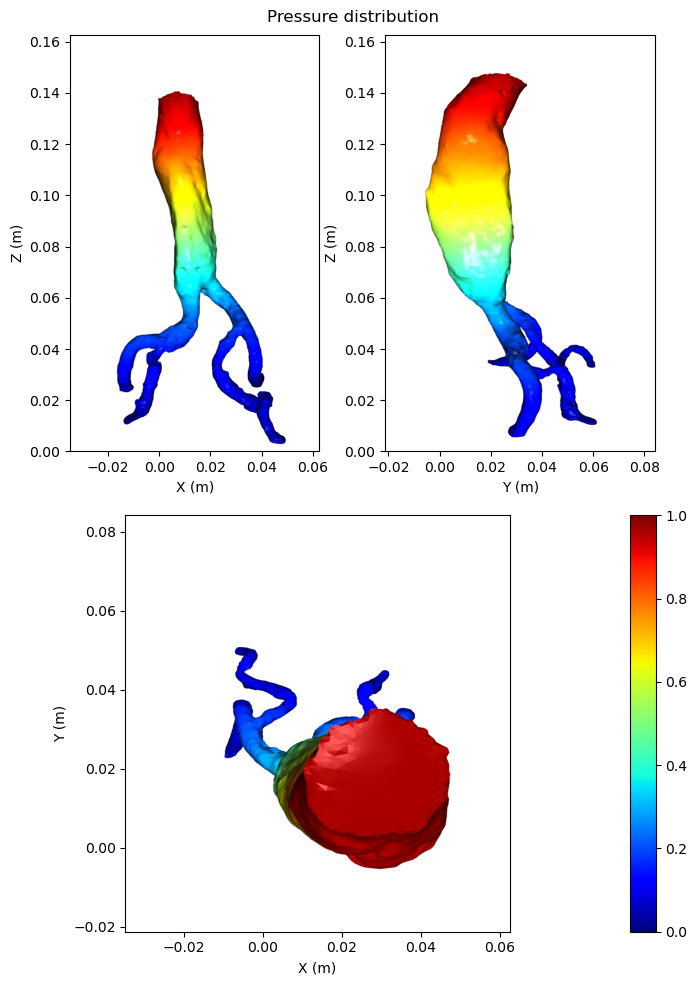}
    \caption{Front, lateral and elevation perspectives of 3D pressure colour map. The colormap represents normalised pressures.}
    \label{fig:wall_pressures}
\end{figure}

The distribution of pressures on the surface of the abdominal aorta is represented in Figure \ref{fig:wall_pressures}, where red colour represents high pressure, green represents mean pressure and blue represents low pressure, as seen in the colour scale. All pressures are scaled in the range of the diastolic and systolic pressure, meaning that blue areas are locally below diastolic pressure.

Although the results are shown here as static images, the actual output of the implemented module is an interactive 3D image of the abdominal aorta with the surface coloured according to the pressure predicted by the model.  

\begin{figure}[H]
    \centering
    \includegraphics[width=0.6\linewidth]{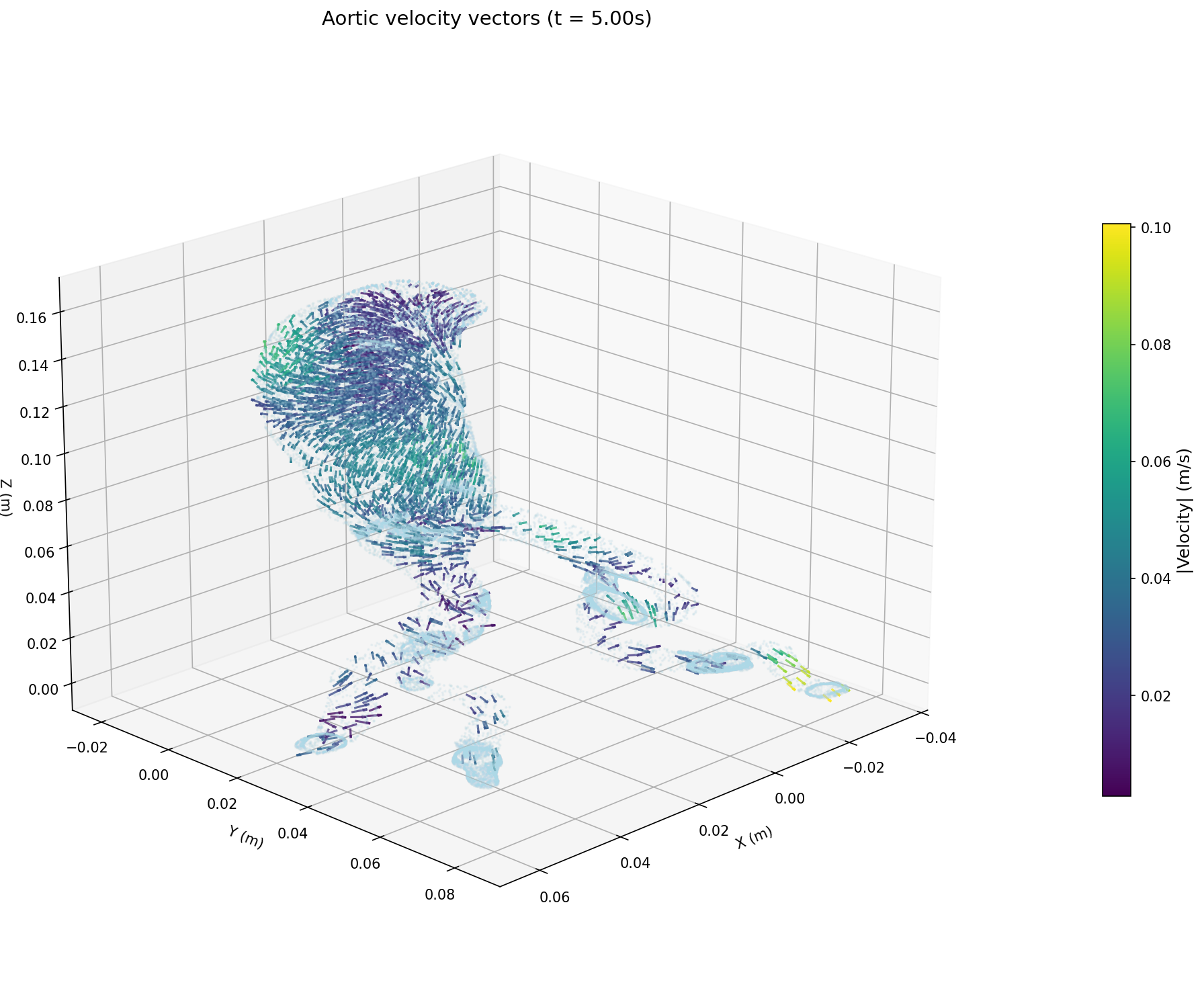}
    \caption{Representation of internal velocity vectors at random domain points for $t=5.0$ s (diastole)}
    \label{fig:streamlines}
\end{figure}

\begin{figure}[H]
    \centering
    \includegraphics[width=0.6\linewidth]{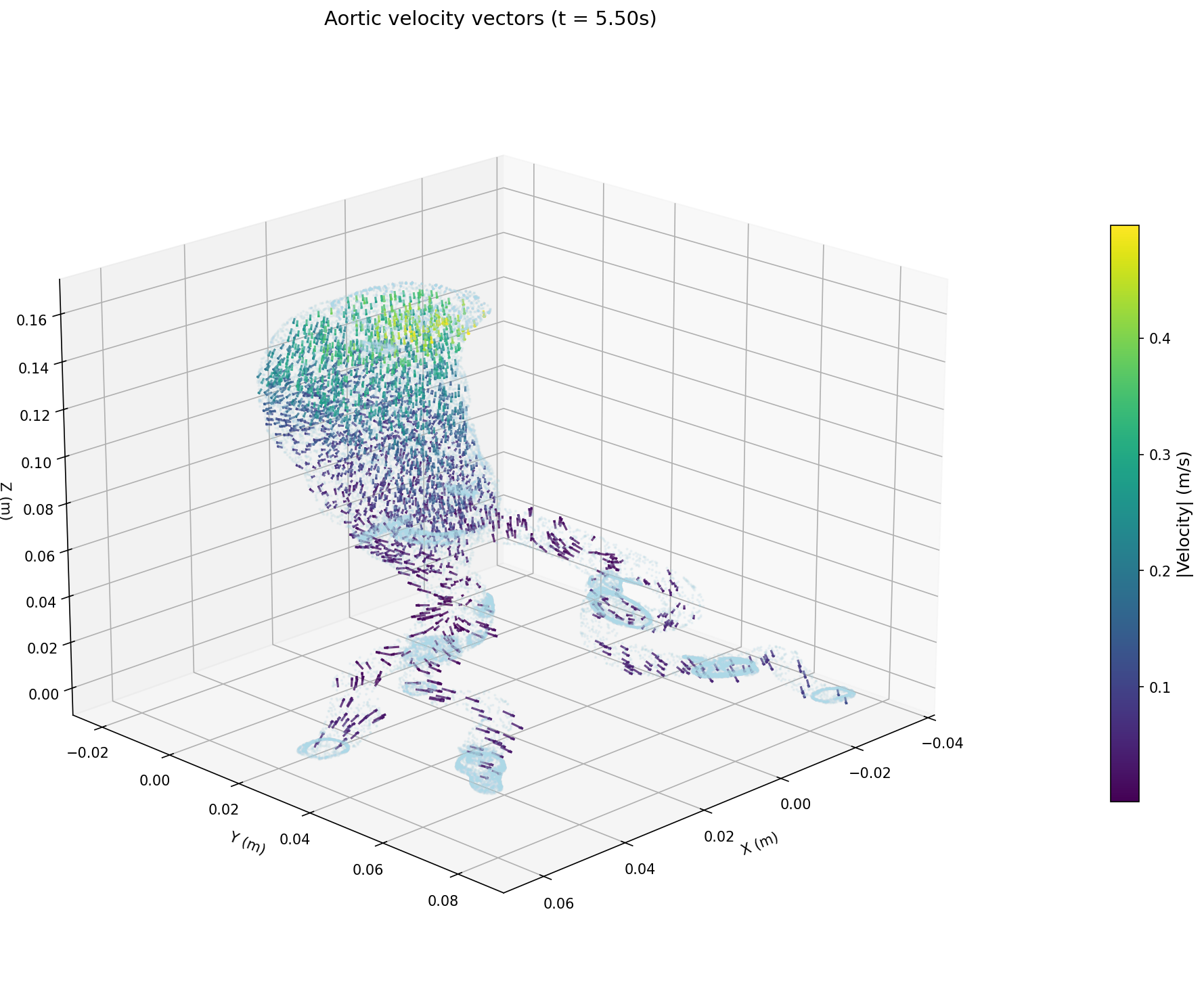}
    \caption{Representation of internal velocity vectors at random domain points for $t=5.5$ s (systole)}
    \label{fig:streamlines_2}
\end{figure}

Figures \ref{fig:streamlines} and \ref{fig:streamlines_2} show the velocity vectors at multiple random points inside the geometry for two different instants of time (the first for the diastole and the second for the systole). This vectors show that peak velocity can be found at the centre of the abdominal aorta, before the common iliac bifurcation. The velocity descends as the blood flows inside the smaller iliac arteries. It can also be seen that there is plenty of direction change, due to collisions with the walls and the appearance of vortices, leading to irregular flux. 

Also, a statistical analysis of the numerical pressure and velocity results was also carried out, separating the individual vector components and subdividing the region of interest into the descending aorta and the iliac bifurcations for two instants of time (diastole and systole). This allowed us to compare the differences in blood‑flow behaviour before and after the aortic bifurcation. The outcomes of this analysis are presented in Tables \ref{tab:stats} and \ref{tab:stats_2}.

\begin{table} [H]
    \centering
    \caption{Statistical results for velocity and pressure in both subsets of the ROI for $t=5.0$ s (diastole).}
    \begin{tabular}{|c|c|c|}
        \hline \textbf{Mean $v_i$ (m/s)} & \textbf{Inlet (Abdom. Aorta)} & \textbf{Outlet (Iliac arteries)} \\
        \hline
        $v_x$ & -0.0004 $\pm$ 0.0121 & 0.0130 $\pm$ 0.0003 \\
        $v_y$ & 0.0102 $\pm$ 0.0073 & 0.0197 $\pm$ 0.0007 \\
        $v_z$ & -0.0077 $\pm$ 0.0134 & -0.0233 $\pm$ 0.0018 \\
        $|v|$ & 0.0232$\pm$0.0025 & 0.0332 $\pm$ 0.0015 \\
        \hline
        \hline \textbf{Pressure (Pa)} & \textbf{Abdom. Aorta} & \textbf{Iliac arteries}  \\
        \hline mean & 9176.08 & 9463.75 \\
        std. dev. & 116.74 & 72.51 \\
        min & 8970.14 & 9245.80 \\
        max & 9466.23 & 9578.57 \\
        \hline
    \end{tabular}
    \label{tab:stats}
\end{table}

\begin{table} [H]
    \centering
    \caption{Statistical results for velocity and pressure in both subsets of the ROI for $t=5.5$ s (systole).}
    \begin{tabular}{|c|c|c|}
         \hline \textbf{Mean $v_i$ (m/s)} & \textbf{Inlet (Abdom. Aorta)} & \textbf{Outlet (Iliac arteries)}  \\
         \hline
        $v_x$ & -0.0213 $\pm$ 0.0032 & -0.0179 $\pm$ 0.0004 \\
        $v_y$ & 0.0148 $\pm$ 0.0237 & 0.0142 $\pm$ 0.0012 \\
        $v_z$ & -0.4819 $\pm$ 0.0597 & -0.0170 $\pm$ 0.0010 \\
        $|v|$ & 0.4832 $\pm$ 0.0597 & 0.0285 $\pm$ 0.0002 \\
        \hline
        \hline \textbf{Pressure (Pa)} & \textbf{Abdom. Aorta} & \textbf{Iliac arteries}  \\
        \hline mean & 13974.50 & 10608.95 \\
        std. dev. & 1122.36 & 782.53 \\
        min & 11761.82 & 9253.69 \\
        max & 16010.73 & 11948.87 \\
        \hline
    \end{tabular}
    \label{tab:stats_2}
\end{table}

As shown in Tables \ref{tab:stats} and \ref{tab:stats_2}, the numerical analysis reveals a clear distinction between the haemodynamic behaviour in the descending abdominal aorta and in the iliac arteries. Mean velocity and pressure values are consistently higher generally in the pre‑bifurcation segment, reflecting the larger vessel calibre and the more coherent axial flow characteristic of this region. In contrast, although the average magnitudes of the lateral velocity components ($v_x$ and $v_y$) decrease after the bifurcation, their standard deviations increase markedly in the iliac arteries. This behaviour is consistent with the more tortuous geometry of the iliac branches, where curvature and branching effects promote greater lateral deviations, wall interactions, and local changes in flow direction.

During diastole ($t = 5.0\,\mathrm{s}$), the mean pressure is relatively uniform throughout the domain, with slightly higher values observed in the iliac arteries. This behavior is consistent with the temporal propagation of the pressure wave and the small pressure gradients typically present during this phase of the cardiac cycle. As for the flow acceleration phase ($t = 5.5\,\mathrm{s}$), a clearly defined pressure gradient develops between the abdominal aorta and the iliac arteries. The pressure is significantly higher upstream and decreases towards the outlets, which is consistent with the expected physiological blood flow from the inlet towards the bifurcations.

Overall, the statistical comparison highlights the expected reduction in axial flow intensity downstream of the bifurcation, together with an increase in lateral velocity variability associated with the more complex post‑bifurcation geometry. Moreover, there is a clear distinction on the results for the diastolic and systolic phases. The ones for the diastole are clearly lower (both for the pressures and the velocities), while the values for the systole are visibly higher. This results are physically consisten with what is expected for the physiological values.

\begin{figure} [H]
    \centering
    \includegraphics[width=1\linewidth]{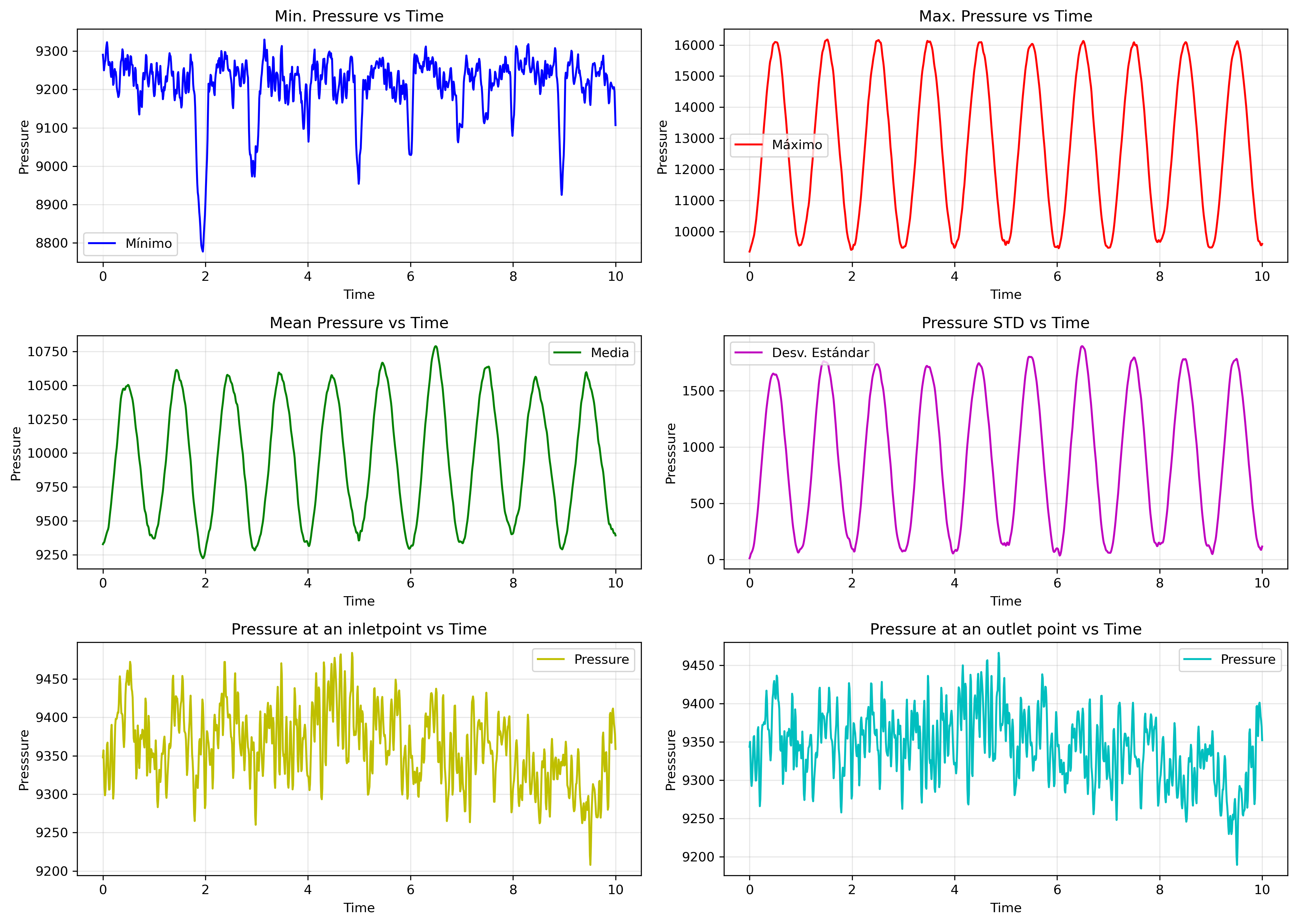}
    \caption{Temporal evolution of the pressure: minimum (top left), maximum (top right), mean (center left), standard deviation (center right), pressure at an inlet point (bottom left), and pressure at an outlet point (bottom right).}
    \label{fig:temporal_evolution}
\end{figure}

Figure \ref{fig:temporal_evolution} represents the temporal evolution of the pressure in our region of interest. Here we can see the evolution of the maximum, minimum, mean, standart deviation and value at the inlet and outlet of the pressure from $t=0$ untill $t=10.0$ s. We can observe here that our model learns the pulsatile behaviour of the blood flow (specifically following the frequency of the inlet pulse).

Finally, the validation of the current model has been performed according to the criteria specified in the previous Section \ref{sec:Validation}, addressing the physical consistency of the model using the corresponding residuals of the continuity and momentum equations to calculate the $L^2$ and $L_{\infty}$ metrics for each, as specified in equations \ref{L2 and Linf} , given the results shown in Table \ref{tab:validacion_pinn}: 

\begin{table}[H] 
    \centering 
    \caption{Validation metrics.}
    \begin{tabular}{|l|c|} 
    \hline \textbf{Metric} & \textbf{Score} \\ 
    \hline Continuity L$^2$ & $3.67 \times 10^{-2}$ \\ 
    Continuity L$\infty$ & $1.08 \times 10^{-1}$ \\ 
    Momentum L$^2$ & $2.8 \times 10^{0}$ \\ 
    Momentum L$\infty$ & $3.60 \times 10^{0}$ \\ 
    \hline 
    \end{tabular} 
     \label{tab:validacion_pinn} 
\end{table}
        
\section{Discussion} \label{sec:Disscusion}

The main obstacle in assessing the results lies in the absence of a CFD based comparative model that could serve as ground truth. Nevertheless, this limitation has been mitigated by analysing the error metrics reported in Table \ref{tab:validacion_pinn} and by comparing both these metrics and the visual outputs of the model with those presented in related studies. Such works include CFD‑ and PINN‑based simulations performed on simpler geometries as well as on vascular structures of comparable complexity.

Beyond validating the physical consistency of the results, as outlined in Section \ref{sec:Validation}, we conducted a comparative analysis with the previously discussed studies, using them as benchmarks for the expected behaviour of blood‑flow simulations in a large‑calibre artery such as the abdominal aorta. Although these works do not address the problem in an identical manner, all reference studies used for comparison rely on similar vascular geometries and use fully three‑dimensional time-dependent formulations.

The pressure distribution obtained by the model is consistent with what has been reported in simulations involving simpler and lower‑dimensional geometries. Specifically, the results align with the expected behaviour in a large‑diameter vessel that branches into smaller arteries, where higher pressures concentrate in the proximal, wider regions and progressively decrease along the bifurcations. This pattern is consistent with observations reported in studies such as Song et al. (2024) \cite{multicasePINN2D} and Kabir et al. (2021)\cite{Num2Dstenosis}.

Building on other studies that address problems of comparable complexity, several works have explored the use of PINNs in 4D flow simulations, yielding results that are consistent with those obtained in the present model. For instance, the pressure profile reported for the abdominal aorta in Zhang et al. (2023) \cite{PINN4Daorta_comp} closely resembles the distribution predicted by our model, aside from the specific geometric characteristics of the aortic segment selected in each case.

Similarly, the study by Cruz González et al. (2025) \cite{PINN_DeepONet_comp} follows a comparable methodological approach, albeit using a simplified vascular geometry of an Abdominal Aorta Aneurysm for their experiments. Their predicted pressure profile exhibits the same qualitative behaviour, with elevated pressures in regions of larger diameter that progressively decrease as the flow moves downstream. In this case, the reported $L^2$ errors are of the same order of magnitude as those obtained in the present work, and in some instances even slightly higher.

\section{Limitations} \label{sec:Limitations}

Although the current model yields satisfactory results, several ways for improvement should be considered. This section addresses the main limitations that were faced during the development of the project, and also the future works leading to solve those known limitations.

One of the main remaining limitations of the present model is the assumption of Newtonian blood behaviour. Although this approximation is commonly adopted in large-vessel haemodynamic simulations, particularly in the aorta where shear rates are generally high, blood exhibits non-Newtonian properties that may influence local flow behaviour under complex haemodynamic conditions \cite{NonNewtonianBlood}. 

In abdominal aortic aneurysms, regions of recirculation, flow separation, and low wall shear stress can amplify these non-Newtonian effects, potentially affecting the accuracy of pressure and velocity predictions near the aneurysmal wall. Incorporating non-Newtonian constitutive models, such as the Carreau-Yasuda or Casson formulations \cite{CarreauYasudaBlood}, into the PINN framework could therefore improve the physical realism of the simulations.

However, extending PINNs to non-Newtonian haemodynamics substantially increases the complexity of the governing equations and the associated optimization process, particularly in three-dimensional and time-dependent domains.

Beyond non-Newtonian fluid dynamics modelling, another important improvement concerns the model’s ability to adapt to different vascular morphologies. At present, the network is trained on a single aortic geometry, and its predictions are therefore limited to that specific configuration, risking to lose much of its accuracy with different aortas. For clinical deployment, however, it would be essential to develop a model capable of generalising to previously unseen geometries, while sacrificing as little accuracy as possible.

Finally, it is worth considering whether more advanced PINN‑based architectures could further enhance performance. Models such as CNN‑PINNs or GCN‑PINNs have demonstrated strong results in other PDE‑based applications, including fluid dynamics \cite{CNN_PINN_Liu, GCN_PINN}. These architectures, however, loses the flexibility of operating directly on point clouds and instead require images or graph structures, which are considerably more demanding in terms of computational resources.
        
\section{Conclusions and future works} \label{sec:Conclusion}

The study has presented a complete workflow for the construction, training, and evaluation of a physics‑informed neural network tailored to simulate haemodynamics in a patient‑specific abdominal aorta. Beyond the formulation of the PINN itself, the work integrates several methodological components that enable the model to operate on realistic vascular geometries. The network was implemented using the DeepXDE library, which provided the computational framework for enforcing the Navier–Stokes equations throughout the domain. To support this, a dedicated preprocessing pipeline was developed to handle the STL‑based anatomical data, extract and classify boundary points, and generate interior sampling points suitable for both training and testing. This pipeline ensured that the model could be trained directly on complex, irregular geometries without the need for mesh generation, while maintaining control over the distribution of points across inlet, outlet, and wall regions.

In parallel, the study incorporated an interactive visualisation system capable of rendering the predicted pressure distribution on the aortic surface in three dimensions. This tool enables dynamic inspection of the haemodynamic fields and provides an intuitive means of assessing the physical plausibility of the results, complementing the quantitative validation metrics. Together, these components form a coherent framework that demonstrates the feasibility of applying PINNs to clinically relevant vascular simulations and lays the groundwork for future extensions involving more complex flow regimes and patient‑specific variability.

In conclusion, a PINN model has been successfully trained to produce the results required for future work aimed at improving the prevention and treatment of abdominal aortic aneurysms. This approach has the potential to reduce the morbidity and mortality associated with such a prevalent condition by providing clinicians with a valuable decision‑support tool that facilitates diagnosis and the anticipation of possible complications.

Looking ahead, several research lines for future development emerge from the limitations identified in this study. A priority for subsequent work is the incorporation of non-Newtonian fluid dynamics into the haemodynamic formulation, as the flow patterns observed in the abdominal aorta suggest the presence of simplified behaviour. Integrating these models, would allow the network to represent more realistic flow regimes, at the cost of increased mathematical and computational complexity.

In addition, enhancing the model’s ability to generalise across diverse vascular morphologies represents an essential step toward clinical applicability. Training the network on multiple aortic geometries, rather than a single configuration, would enable more robust predictions in patient‑specific scenarios. Finally, exploring more advanced PINN‑based architectures, such as CNN‑PINNs or GCN‑PINNs, may offer further improvements in accuracy for complex PDE‑driven problems, though these approaches require more structured data and greater computational resources as well. Together, these directions outline a clear path for strengthening the model’s predictive capabilities and expanding its potential for real‑world clinical integration.

\section{Ethics Statement}
All procedures performed in this study were conducted in full compliance with the relevant national legislation and institutional guidelines of the University of Alicante and Avamed Synergy.

Informed consent for the acquisition and research use of imaging data was obtained from all human subjects by the clinical provider (Avamed Synergy) prior to data anonymisation and transfer. All personal identifiers were removed before data processing, and the privacy rights of all participants were strictly protected throughout the study. No additional interventions, experiments, or procedures were performed on human subjects specifically for this research.

\newpage

\bibliographystyle{ieeetr}
\bibliography{cas-refs}

\end{document}